\documentclass[twoside,fleqn]{article}
\usepackage{espcrc2}
\usepackage{psfig}

\title{Excited baryons from Bayesian priors and overlap fermions
\thanks{presented by F.X. Lee at Lattice 2002.}
\thanks{This work is supported in part by U.S. Department of Energy 
under grants DE-FG05-84ER40154 and DE-FG02-95ER40907.}}

\author{F.X.~Lee\address[GW]{Center for Nuclear Studies,
        George Washington University, Washington, DC 20052, USA}
        \address[JL]{Jefferson Lab, 12000 Jefferson Avenue, Newport News, VA 23606, USA},
        S.J.~Dong\address[UK]{Department of Physics \& Astronomy,
        University of Kentucky, Lexington, KY 40506, USA},
        T.~Draper\addressmark[UK],
        I.~Horv\'ath\addressmark[UK],
        K.F.~Liu\addressmark[UK],
        N.~Mathur\addressmark[UK],
        J.B.~Zhang\address{CSSM, University of Adelaide, Adelaide, SA 5005, Australia}}

\begin{document}

\begin{abstract}
Using the constrained-fitting method based on Bayesian priors, 
we extract the masses of the two lowest states of octet and decuplet baryons
with both parities.
The calculation is done on quenched $16^3\times 28$ lattices of $a = 0.2$ fm using 
an improved gauge action and overlap fermions, with the pion mass as low as 180 MeV.
The Roper state $N(1440)\frac{1}{2}^+$ is clearly observed for the first time 
as the 1st-excited state of the nucleon from the standard interpolating field. 
Together with other baryons, our preliminary results indicate 
that the level-ordering of the 
low-lying baryon states on the lattice is largely consistent with experiment. 
The realization is helped by cross-overs between the excited $\frac{1}{2}^+$ 
and $\frac{1}{2}^-$ states in the region of $m_\pi \sim$ 300 to 400 MeV.
\end{abstract}

\maketitle

The rich structure of the excited baryon spectrum, as documented by the 
particle data group~\cite{pdg00},
provides a fertile ground for exploring
the nature of quark-quark interactions.
One outstanding example is the ordering of the lowest-lying states
which has the order of positive and negative-parity excitations inverted between 
$N$, $\Delta$ and $\Lambda$ channels. 
%
%
Conventional quark models have difficulty explaining 
the ordering in a consistent manner.
There are two contrasting views.
One is from the constituent quark model~\cite{Isgur78,Capstick86}
which has the interaction dominated by one-gluon-change type, {\it i.e.}, 
color-spin $\lambda^c_1\cdot \lambda^c_2 \vec{\sigma}_1 \cdot \vec{\sigma}_2$.
The other is based on Goldstone-boson-exchange~\cite{Glozman96}
which has flavor-color 
$\lambda^f_1\cdot \lambda^f_2 \vec{\sigma}_1 \cdot \vec{\sigma}_2$ as the dominant part.
Even though evidence from valence QCD~\cite{Liu99} 
supports the flavor-color picture,
the challenge of reproducing the ordering still faces lattice QCD.

There exist a number of lattice studies of the
excited baryon spectrum using a variety of
actions~\cite{Derek95,Lee,Wally02,sas00,dgr00}.
The nucleon channel is the most-studied, focusing on
two independent local fields:
\begin{equation}
\chi_1 = \epsilon^{abc}\left ( u^{Ta} C \gamma_5 d^b \right ) u^c,
\end{equation}
\begin{equation}
\chi_2 = \epsilon^{abc} \left ( u^{Ta} C d^b \right ) \gamma_5 u^c.
\end{equation}
\vspace*{-0.3cm}
$\chi_1$ is the standard nucleon operator, while $\chi_2$, which has 
a vanishing non-relativistic limit, is sometimes 
referred to as the `bad' nucleon operator.
Note that baryon interpolating fields couple to both positive and negative-parity states,
which can be separated by well-established parity-projection techniques.
The consensus so far appears to be that, first, the negative-parity 
splitting of $N(\frac{1}{2}^-)$ is largely established and consistent with experiment.
Secondly, the Roper state $N^\prime(\frac{1}{2}^+)$ as the 1st-excited state
of the nucleon is still elusive.
Since $\chi_2$ couples little to the nucleon ground state, there was
initial speculation that it couples to the Roper state.
But that identification has been mostly abandoned since the mass extracted 
from $\chi_2$ is consistently too high.
What $\chi_2$ couples to remains an open question.

\begin{figure}
 \centerline{\psfig{file=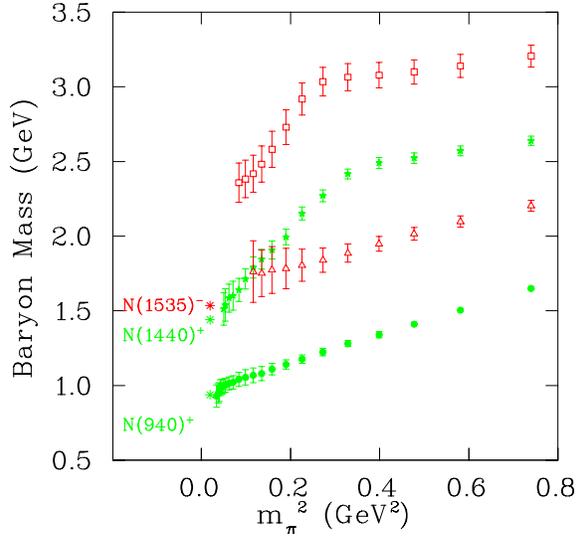,width=7.5cm,angle=0}}
\vspace*{-1.0cm}
\caption{Solid symbols denote $N(\frac{1}{2}^+)$ states: 
ground ($\bullet$) and 1st-excited ($\star$).
Empty symbols denote $N(\frac{1}{2}^-)$ states: 
lowest ($\triangle$) and 2nd lowest (\fbox{}).
The experimental points ($*$) are taken from PDG~\protect\cite{pdg00}.}
\label{mass_nuc_dong}
\end{figure}
\begin{figure}
 \centerline{\psfig{file=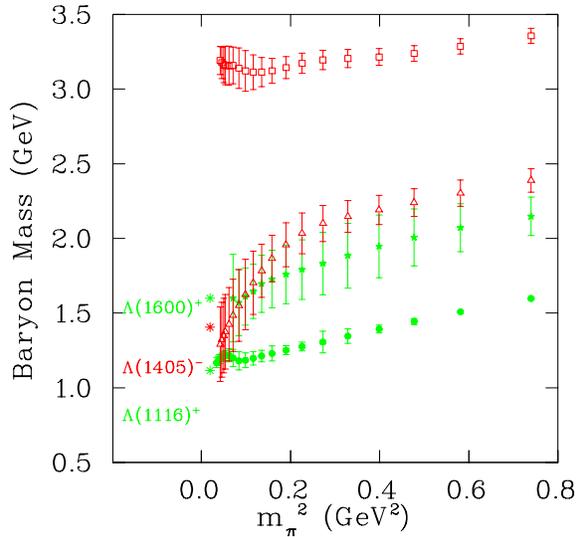,width=7.5cm,angle=0}}
\vspace*{-1.0cm}
\caption{Similar to Fig.~\protect\ref{mass_nuc_dong}, 
but for $\Lambda(\frac{1}{2}^\pm)$ states.}
\label{mass_lam}
\end{figure}
\begin{figure}
 \centerline{\psfig{file=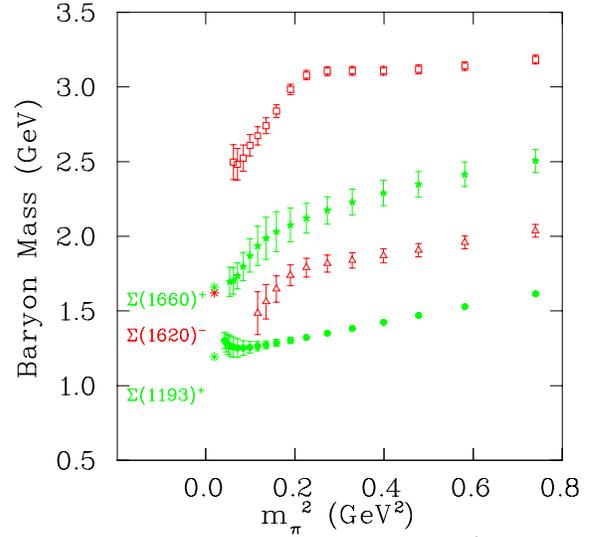,width=7.5cm,angle=0}}
\vspace*{-1.0cm}
\caption{Similar to Fig.~\protect\ref{mass_nuc_dong}, 
but for $\Sigma(\frac{1}{2}^\pm)$ states.}
\label{mass_osig}
\end{figure}
\begin{figure}
 \centerline{\psfig{file=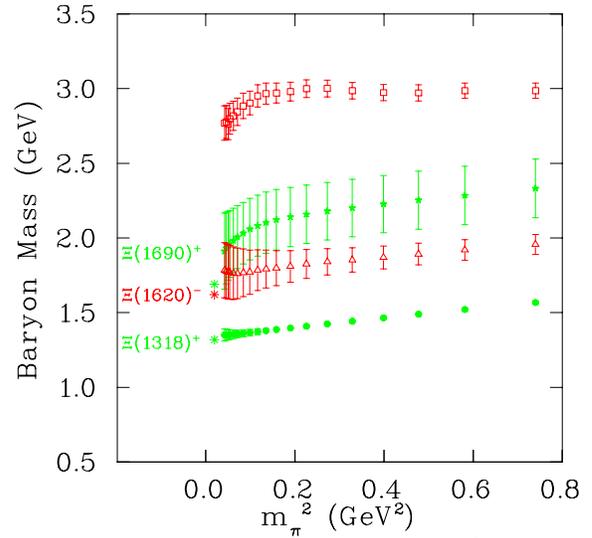,width=7.5cm,angle=0}}
\vspace*{-1.0cm}
\caption{Similar to Fig.~\protect\ref{mass_nuc_dong}, 
but for $\Xi(\frac{1}{2}^\pm)$ states.}
\label{mass_oxi}
\end{figure}
\begin{figure}
 \centerline{\psfig{file=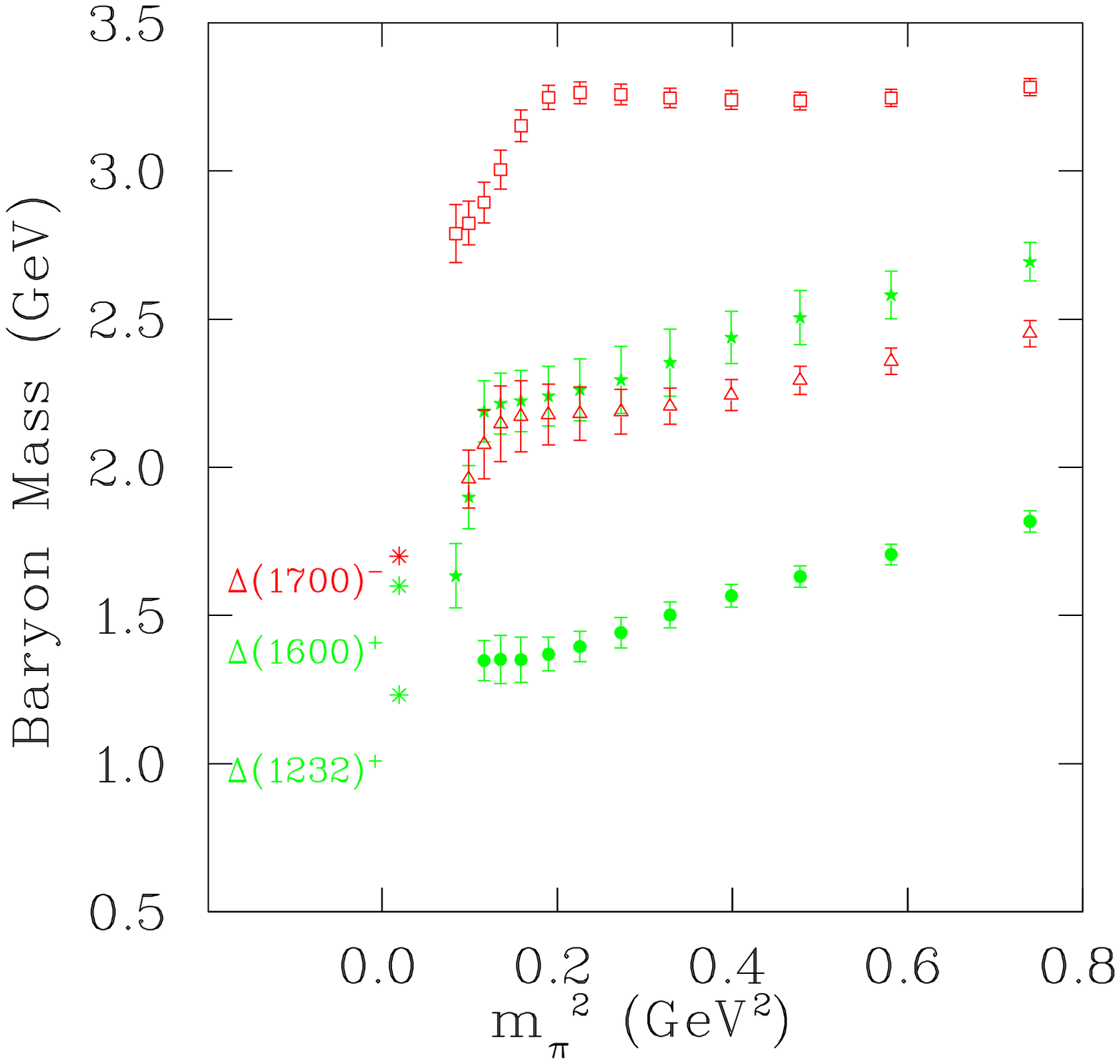,width=7.5cm,angle=0}}
\vspace*{-1.0cm}
\caption{Similar to Fig.~\protect\ref{mass_nuc_dong}, 
but for $\Delta(\frac{3}{2}^\pm)$ states.}
\label{mass_del}
\vspace*{-1.0cm}
\end{figure}

The lattice size we use is $16^3\times 28$ 
with the scale of $a=0.202(1)$ fm set from $f_\pi$, 
which is our preferred choice for scale~\cite{Dong01}.
We consider a wide range of quark masses: 
26 masses with the lowest $m_\pi=180$ MeV (or $m_\pi/m_\rho$=0.248), very close 
to the physical limit,
and with 18 masses below the strange quark mass.
We analyzed 80 configurations.
Details of the simulation can be found in~\cite{Draper02}.

We adapted the constrained curve-fitting method advocated in~\cite{Lepage02,Morn02},
adhering to the following guidelines: 
a) fit as many time slices in the correlation function $G_{\rm data}(t)$ and 
as many terms in $G_{\rm theory}(t)$ as possible;
b) use prior knowledge, such as $A>0$ and $E_n-E_{n-1}>0$; 
c) seek guide for priors from a subset of data (empirical Bayes method);
d) un-constrain the term of interest in $G_{\rm theory}(t)$ 
to have conservative error bars.
The details of the implementation are discussed in~\cite{Dong02}.

Fig.~\ref{mass_nuc_dong} to Fig.~\ref{mass_del} show the results in various channels.
To emphasize the small mass region, only the masses starting from the strange quark mass are shown.
The most significant feature is that the ordering among $N(938)\frac{1}{2}^+$, 
$N^\prime(1440)\frac{1}{2}^+$,
$N(1535)\frac{1}{2}^-$, and $\Lambda(1405)\frac{1}{2}^-$ is consistent with experiment, 
which is the first time this has been seen on the lattice.
It comes about with 
cross-overs between the excited $\frac{1}{2}^+$ and $\frac{1}{2}^-$ states in 
the region of $m_\pi \sim$ 300 to 400 MeV.

Further study is under way to check the stability of the fitting algorithm, especially in the small mass region where the signal worsens,  and to automate the fitting process.
To make sure that the results are not due to finite-volume effects~\cite{Morn02a} 
(our lowest $Lm_\pi \approx 3$), 
we are repeating 
the entire calculation on a smaller lattice of $12^3\times 28$ with all other parameters 
fixed. The lowest $m_\pi$ on this lattice is about 250 MeV, small enough 
to probe the crossover region.

\end{document}